\documentclass[showpacs,preprintnumbers,amsmath,amssymb,aps]{revtex4}

\usepackage{graphicx}
\usepackage{dcolumn}
\usepackage{bm}
\newcommand{\bmat}{\left(\begin{array}}
\newcommand{\emat}{\end{array}\right)}
\newcommand{\be}{\begin{equation}}
\newcommand{\ee}{\end{equation}}
\newcommand{\bea}{\begin{eqnarray}}
\newcommand{\eea}{\end{eqnarray}}
\newcommand{\ba}{\begin{eqnarray}}
\newcommand{\ea}{\end{eqnarray}}
\newcommand{\nn}{\nonumber}

\begin{document}

\preprint{}

\title{Is there a paradox in $CP$ asymmetries of $\tau^{\pm} \to
K_{L,S}\pi^{\pm}\nu$ decays?}

\author{G. Calder\'on}
\email{german.calderon@mail.uadec.mx}
\affiliation{Facultad de
Ciencias F\' isico Matem\'aticas, UAdeC, 25280 Saltillo, Coahuila,
M\'exico}
\author{D. Delepine}
 \email{delepine@fisica.ugto.mx}
 \affiliation{Instituto de F\'isica de la Universidad de Guanajuato,
 C.P. 37150, Le\'on, Guanajuato, M\'exico.}%

\author{G. L\'opez Castro}
\email{glopez@fis.cinvestav.mx}
\affiliation{ Departamento de F\'\i
sica, Cinvestav, A.P. 14-740, 07000 M\'exico D.F, M\'exico}

\date{\today}

\begin{abstract}
Based on the description of unstable $K_{L,S}$ particles in quantum
field theory (QFT), we compute the time-dependent probabilities for
transitions between asymptotic states  in $\tau^{\pm} \to
[\pi^+\pi^-]_K\pi^{\pm}\nu$ decays, where the pair $[\pi^+\pi^-]_K$
is the product of (intermediate state) neutral kaon decays. Then we
propose a  definition of $\tau$ decays into $K_L$ and $K_S$ states,
which reflects into the cancellation between their CP rate
asymmetries, thus solving in a natural way the paradox pointed out
in Ref. \cite{Bigi:2005ts}. Since our definition of $K_{L,S}$ final
states in $\tau$ decays is motivated on experimental grounds, our
predictions for the integrated CP rate asymmetries can be tested in
a dedicated experiment.
\end{abstract}

\maketitle

\section{Introduction}

 In a recent paper, Bigi and Sanda \cite{Bigi:2005ts} have pointed out
that $\tau^{\pm} \to K_{L,S}\pi^{\pm}\nu_{\tau}$ decays exhibit a CP
asymmetry  of the same size as the one measured in the charge
asymmetry of semileptonic  $K_L$ decays. The `known' CP rate
asymmetries for $K_S$ and $K_L$ final states \cite{Bigi:2005ts}
\ba
\frac{\Gamma(\tau^+\to K_S\pi^+\bar{\nu}) - \Gamma(\tau^-\to
K_S\pi^-\nu)}{\Gamma(\tau^+\to K_S\pi^+\bar{\nu}) + \Gamma(\tau^-\to
K_S\pi^-\nu)} &=& |p|^2-|q|^2\ , \\
\frac{\Gamma(\tau^+\to K_L\pi^+\bar{\nu}) - \Gamma(\tau^-\to
K_L\pi^-\nu)}{\Gamma(\tau^+\to K_L\pi^+\bar{\nu}) + \Gamma(\tau^-\to
K_L\pi^-\nu)} &=& |p|^2-|q|^2 \ ,
\ea
turn out to be identical. If true, this would indicate a paradox because
the total rates of $\tau^+$ and $\tau^-$ would be different, in
contradiction with the CPT theorem.

   Bigi and Sanda \cite{Bigi:2005ts} proposed a solution to this
contradiction  by looking  at the time evolution  of the $K^0$
($\bar{K}^0$) state produced in  $\tau^+$
($\tau^-$) decays at $t=0$. They concluded that the sum of
time  integrated rates (from $t=0$ to $\infty$) over all the final states
that can be reached by neutral kaon decays  is free from  such CP
asymmetries, restoring the equality of the $\tau^{\pm}$ lifetimes. In
their discussion of the problem \cite{Bigi:2005ts}, the interference of
the $K_{S,L}$  states in the time-dependent rates plays  an essential role
in the cancellation of  the contributions of pure $K_L$ and $K_S$ exponential decays.

   In this paper we approach this problem from a different point of view.
Based on the description of unstable $K_{L,S}$ particles in quantum
field theory (QFT), we compute the time evolution of transition
amplitudes between physical ({\it in} and {\it out}) states. Let us
point out that, in the evaluation of the matrix elements giving rise
to  the rates that enter Eqs. (1,2), neutral kaons ($K_{L,S}$) are
assumed to be {\it asymptotic} physical states (defined as {\it
outgoing} states at $t\to +\infty$). Since unstable particles in QFT
enter as intermediate states of the physical amplitudes, the paradox
contained in Eqs. (1,2) does not appear. We compute the
time-dependent probabilities for transitions between asymptotic
states in $\tau^{\pm} \to [\pi^+\pi^-]_K\pi^{\pm}\nu$, where
$[\pi^+\pi^-]_K$ is the pair produced from neutral kaon decays. Then
we propose a definition of $\tau$ decays into $K_{L,S}$ states,
which reflects into  a natural cancellation between the  CP rate
asymmetries defined in Eqs. (1) and (2).

Searches for CP violation effects in a double kinematical distribution
of $\tau^{\pm} \to K_S \pi^{\pm}\nu_{\tau}$ decays have been pursued
recently by the CLEO Collaboration \cite{Bonvicini:2001xz}. Prospects
for improved  experimental searches are interesting in the light of the
larger data  samples of $\tau$ pairs accumulated at $B$-factories
\cite{igonkina}. These
exclusive decays can be used to provide further tests on the
violation of the CP symmetry
\cite{Bigi:2005ts,Bonvicini:2001xz,Kuhn:1996dv,Delepine:2005tw}.
On the other hand, within the standard model, the CP rate asymmetry turns
out to be negligibly small (of order $10^{-12}$) in $\tau^{\pm} \to
K^{\pm} \pi^0\nu_{\tau}$ decays \cite{Delepine:2005tw}, opening a large
window to consider the effects of New Physics contributions.
Indeed,  virtual effects of supersymmetric particles may enhance this  CP
rate asymmetry to the level of $10^{-7}\sim 10^{-6}$
\cite{Delepine:2006fv}.

\section{Time-dependent amplitudes in $\tau$ lepton decays}
 In this section we focus on calculation of the time evolution
amplitudes of $\tau$ lepton decays into asymptotic physical states, as
dictated by the S-matrix formalism of QFT.

\begin{figure}
\includegraphics[width=12cm]{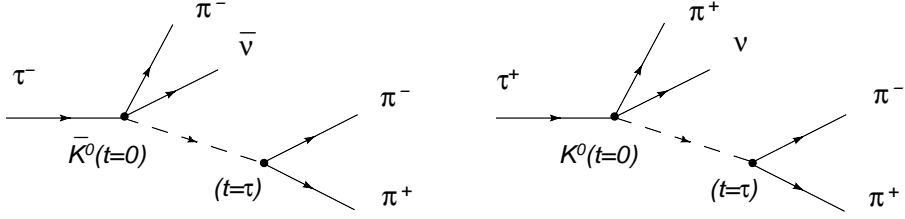}\\
\caption{$\tau$ lepton decay into neutral kaon states of definite flavor
at time $t=0$. These unstable kaon states evolve as a mixing of two
orthogonal states of definite mass and width $(K_L$ and $K_S)$ and decay
into a $\pi^{+} \pi^{-}$ pair at time $t=\tau$.}
\end{figure}

Let us start by defining, at time $t=0$, the $\tau$ lepton decays
into virtual strange states of definite flavor  ($\tau^+ \to K^0
\pi^+ \nu$ or $\tau^-  \to \bar K^0 \pi^-  \bar \nu )$. The decay
amplitudes of these two processes are the same owing to the CPT
invariance  of weak  interactions, and will be denoted by $A$. After
being produced, initial kaon states of definite flavor evolve as a
mixing of two {\it orthogonal} states of definite mass and width
($K_L$ and $K_S$) and decay a later time $t=\tau$ into a final state
$f$. For definiteness we will confine ourselves to $f=\pi^+\pi^-$
(see Fig. 1).

 In order to introduce the CP violating effects, we change from the basis
of definite flavor $(K^0, \bar{K}^0)$ to the basis of definite CP parity
$(K_1,\ K_2)$ of neutral kaons:
\be
\left(\begin{array}{c} K_1 \\ K_2 \end{array}\right) \ \equiv \
\frac{1}{\sqrt{2}}\left(\begin{array}{ccc}
1 & 1 \\
1 & -1
\end{array}\right)\ \left(\begin{array}{c}
K^0 \\ \bar{K^0}
\end{array}\right)
\ee

\noindent with the convention $CP|K^0 \rangle = |\bar K^0 \rangle$.

In momentum space, the transition amplitudes for the $K_1- K_2$
system propagating with momentum $p$ from $t=0$ to $t=\tau$, are given by
(see for example \cite{beuthe}):
\be
 D^{K_1-K_2}_R(p^2) = \frac{1}{1-\epsilon^2}
\left(\begin{array}{cc}
1 & \epsilon\\
\epsilon & 1
\end{array}\right)
\left(\begin{array}{cc}
d^{-1}_S & 0\\
0 & d^{-1}_L
\end{array}\right)
\left(\begin{array}{cc}
1 & -\epsilon\\
-\epsilon & 1
\end{array}\right)\ ,
\ee
where $d_{S,L}\equiv p^2-M_{S,L}^2+iM_{S,L}\Gamma_{S,L}$. Equivalently,
the matrix $D_D(p^2)=\mbox{\rm diag}(d^{-1}_S,d_{L}^{-1})$ describes the
propagation of the two independent physical modes of definite mass
($M_{L,S}$) and width ($\Gamma_{L,S}$).

Following the procedure discussed in Ref. \cite{beuthe}, we compute now
the time-dependent transition amplitude of the whole $\tau^-(q)
\to \bar{K}^0(p_1+p_2; t=0)\pi^-(k) \nu(q') \to
\pi^+(p_1)\pi^-(p_2)\pi^-(k)\nu(q')$ decay process. The pair
$\pi^+(p_1)\pi^-(p_2)$ is the product of kaon decay at time $t=\tau$,
thus we define $E=p^0_1 + p^0_2$ as the total energy of the $\pi^+ \pi^-$
pair and  $E_{S,L}=
\sqrt{(\overrightarrow{p_1}+\overrightarrow{p_2})^2+m^2_{S,L}}$. The
time-dependent amplitude becomes:
\ba {\cal T}_-(\tau) &=& (2\pi)^4 \; \delta^{(4)}(q-k-q'-p_1-p_2)
\frac{A}{1-\epsilon}\;{\cal M}(K_1\to \pi^+\pi^-)\; e^{iE\tau}\nn\\
&& \times\left\{\frac{1+\chi_{+-}\ \epsilon}{2E_S} \; e^{-iE_S \tau}
\; e^{-\frac{1}{2} \Gamma_S \frac{m_S}{E_S}\tau}
 - \frac{\epsilon + \chi_{+-}}{2E_L} \; e^{-iE_L \tau} \;
e^{-\frac{1}{2} \Gamma_L \frac{m_L}{E_L} \tau} \right \}\ .
\ea
Similarly, we can obtain the corresponding time-dependent amplitude for
the  $\tau^+(q)\to K^0(p_1+p_2; t=0)\pi^+(k) \bar{\nu}(q') \to
\pi^+(p_1)\pi^-(p_2)\pi^+(k)\bar{\nu}(q')$ decay:
\ba {\cal T}_+(\tau) &=& (2\pi)^4 \; \delta^{(4)}(q-k-q'-p_1-p_2)
\frac{A}{1+\epsilon}\;{\cal M}(K_1\to \pi^+\pi^-)\; e^{iE\tau}\nn\\
&& \times\left\{\frac{1+\chi_{+-}\ \epsilon}{2E_S} \; e^{-iE_S\tau}
\; e^{-\frac{1}{2} \Gamma_S \frac{m_S}{E_S} \tau}
 + \frac{\epsilon + \chi_{+-}}{2E_L} \; e^{-iE_L\tau} \; e^{-\frac{1}{2}
\Gamma_L \frac{m_L}{E_L} \tau} \right\}\ . \ea

In the above expressions we have defined the direct CP
violation parameter \cite{beuthe}

\be \chi_{+-} \equiv \frac{{\cal M} (K_2 \to \pi^+\pi^-)}{{\cal M}
(K_1 \to \pi^+\pi^-)}\ , \ee and ${\cal M}(K_{1,2}\to \pi^+\pi^-)$
denote the `instantaneous' decay amplitudes (at $t=\tau$) of the CP
eigenstates into a $\pi^+\pi^-$ pair. Note that in Eqs. (5,6) we do
not need to symmetrize the decay amplitudes  because identical pions
are produced at different space-time locations.

For further use in the following discussion we introduce  the usual
CP violation parameter \ba \eta_{+-} &=& \frac{\epsilon +
\chi_{+-}}{1+\chi_{+-}\ \epsilon}= |\eta_{+-}|e^{i\phi_{+-}}\ .\ea
The time-dependent amplitudes given in Eqs. (5,6) are defined in an
arbitrary reference frame \cite{beuthe}. Now, we choose the center
of mass frame of the pion pair produced at $t=\tau$, which means
$E_{S,L}=M_{S,L}$. The expressions for the time-dependent
probabilities (up to the second order in  the CP violation parameter
and neglecting direct $CP$ violation, {\it i.e.} $\chi_{+-}=0$) are
given by: \ba |{\cal T}_-(\tau)|^2 &\simeq&
\frac{B(1+2Re[\epsilon])}{4M^2_S} [e^{-\Gamma_S \tau} + |\epsilon|^2
e^{-\Gamma_L \tau} -
2|\epsilon|e^{-\frac{1}{2}(\Gamma_S+\Gamma_L)\tau}\cos(\Delta m
\tau-\phi_{+-})]\ea

\ba |{\cal T}_+(\tau)|^2 &\simeq&
\frac{B(1-2Re[\epsilon])}{4M^2_S}[e^{-\Gamma_S \tau}+
|\epsilon|^2e^{-\Gamma_L \tau} +
2|\epsilon|e^{-\frac{1}{2}(\Gamma_S+\Gamma_L)\tau}\cos(\Delta m
t-\phi_{+-})],
\ea
where we have defined the common factor $B = (2\pi)^4\
\delta^{(4)}(q+q'-k-k'-p_1-p_2) \ |A {\cal M} (K_1 \to \pi^+\pi^-)
\; e^{iE\tau}|^2$ and $\Delta m=M_L-M_S$. Such
expressions for the time-dependent probabilities are similar to the ones
obtained in the framework of the usual Lee-Oehme-Yang formalism
\cite{ww} and used in the data analysis of the CPLEAR experiment
\cite{cplear1}.

With the above expressions we can calculate the time-dependent CP
rate asymmetry for the processes under consideration. We get the
result:

\ba A_{+-}(\tau) &=&  \frac{|{\cal T}_-(\tau)|^2 - |{\cal
T}_+(\tau)|^2} {|{\cal T}_-(\tau)|^2 + |{\cal T}_+(\tau)|^2}\nn\\
&\simeq& 2 Re[\epsilon] \left[\frac{-\frac{1}{\cos
\phi_{+-}}e^{-1/2(\Gamma_S+\Gamma_L)\tau}\cos(\Delta m
\tau-\phi_{+-})+e^{-\Gamma_S \tau}+|\epsilon|^2e^{-\Gamma_L
\tau}}{e^{-\Gamma_S \tau}+ |\epsilon|^2e^{-\Gamma_L
\tau}}\right].\label{cpasymmetry}\ea
Note that this CP asymmetry, defined for an arbitrary decay time $\tau$
of neutral kaons, does not vanish.

\section{ A definition for the $K_{L,S}$ states}

 Strictly speaking, within the S-matrix formalism we can not define
physical amplitudes for final states containing unstable particles (they
are not asymptotic states). Instead, unstable particles appear as
intermediate states of the S-matrix amplitude connecting the
production and decay processes. However, we can resort to a definition of
$\tau^{\pm} \to K_{L,S}\pi^{\pm}\nu$ decay rates by introducing a time
scale $T$ which allows to separate neutral kaon decays that occur at short
and long decay times.

Thus, in order to define the $CP$ asymmetry for $\tau$ decays into an
specific $K$ meson final state, one has first to adopt a definition of the
amplitude with $K_{L,S}$ states. Due to the large difference between $K_L$
($t_L$) and  $K_S$ ($t_S$) lifetimes, it is possible to adopt a
procedure to separate `$K_L$' and `$K_S$' events. We introduce a time
scale $T$
such that $t_S \ll T \ll t_L$. In this way, decays of neutral kaons that
occur for $t \sim O( t_S)$ can be identified as $K_S$ events, while those
taking place for $t \sim O(t_L)$ would be identified as $K_L$ events.
As is usually done,  to talk about a beam of $K_L$ mesons we
should look at $\pi^+\pi^-$ pairs produced  after a time $t \geq T$.
Inversely, at short decay times ($t \leq T$), our $\pi^+\pi^-$ pairs
will originate  mainly from decays of the $K_S$ component.

Thus, in  analogy with  Ref. \cite{Bigi:2005ts}, we can define the
$CP$ rate asymmetry of $\tau$ leptons into $K_{S,L}\pi$ final states
from our Eq. (\ref{cpasymmetry}). The CP rate asymmetries for $\tau$
decays into $K_S$ (respectively $K_L$) become:

\ba A_{CP}^S& \approx &
\frac{\int_0^{T}\left(-2|\epsilon|e^{-1/2(\Gamma_S+\Gamma_L)\tau}\cos(\Delta
m \tau-\phi_{+-}) +2 Re[\epsilon]e^{-\Gamma_S
\tau}\right)d\tau}{\int_0^{T}e^{-\Gamma_S \tau}d\tau}\ , \\
A_{CP}^L& \approx & \frac{\int_T^{\infty}2Re[\epsilon]e^{-\Gamma_L
\tau}d\tau}{\int_T^{\infty}e^{-\Gamma_L \tau}d\tau}\ , \ea where we
have used the approximations $\Gamma_S \gg \Gamma_L$ and $t_S \ll T
\ll t_L$. A straightforward evaluation of the CP asymmetries from
Eqs. (11,12) gives:
\begin{equation}
A_{CP}^L\approx - A_{CP}^S \approx 2 Re[\epsilon]\ .
\end{equation}
In other words, the CP asymmetry integrated over all decay times for
a {\it given} physical final states does not exhibit the paradox
discussed in the Introduction.  Furthermore, it comes out that the
`experimentally motivated' definitions  of the CP asymmetries for
$\tau^{\pm}$ decays into $K_L$ and $K_S$ proposed above cancel each
other.

\section{Conclusions}

In the framework of the S-matrix formalism of QFT, we compute the
time evolution amplitudes for $\tau^{\pm} \to  [\pi^+\pi^-]_{\cal
K}\pi^{\pm}\nu_{\tau}$ decays, where $[\pi^+\pi^-]_{\cal K}$ denote
the decay products of a unstable neutral kaon. By introducing a
definition of decay rates for short and long decay times of neutral
unstable kaons, we show that the integrated CP rate asymmetries of
these decays cancels each other in a natural way. Therefore, in the
description of unstable states based on the S-matrix formalism there
is not a paradox like the one described in Ref. \cite{Bigi:2005ts}.
Since our definition of $K_{L,S}$ final states in $\tau$ decays is
motivated on experimental grounds, our predictions for the
integrated CP rate asymmetries (Eqs. (14)) can be tested in a
dedicated experiment.

\acknowledgments{G.C. and D.D. want to thank the hospitality of
G.L.C. at the Physics Department of Cinvestav where part of this work was
done. The work of D.D. was supported by Conacyt (M\'exico) grants 46195
and PROMEP grant UGTO-PTC. G.L.C. acknowledges the financial support from
Conacyt.}

\end{document}